\documentclass[prb,aps,twocolumn,superscriptaddress,showpacs,floatfix]{revtex4}

\usepackage{amsmath,amssymb}
\usepackage{graphicx}
\usepackage{version}
\DeclareGraphicsExtensions{.eps,.ps}

\begin{document}

\title
{Single-particle and Interaction Effects on the Cohesion and
Transport and Magnetic Properties of Metal Nanowires at Finite
Voltages}
\author{C.-H. Zhang}
\affiliation{Department of Physics, University of Arizona,
1118 E. 4th Street, Tucson, AZ 85721}
\affiliation{Department of Physics, Indiana University,
727 E. 3th Street, Bloomington, IN 47405}
\date{Submitted: \today}

\begin{abstract}
%%%%%%%%%%%%%%%%%%%%%%%%%%%%%%%%%%%%%%%%%%%%%%%%%%%%%%%%%%%%%%%%%%%%%%%%%%%%%%%
The single-particle and interaction effects on the cohesion,
electronic transport, and some magnetic properties of metallic
nanocylinders have been studied at finite voltages by using a
generalized mean-field electron model. The electron-electron
interactions are treated in the self-consistent Hartree
approximation. Our results show the single-particle effect is
dominant in the cohesive force, while the nonzero magnetoconductance
and magnetotension coefficients are attributed to the interaction
effect. Both single-particle and interaction effects are important
to the differential conductance and magnetic susceptibility.
%%%%%%%%%%%%%%%%%%%%%%%%%%%%%%%%%%%%%%%%%%%%%%%%%%%%%%%%%%%%%%%%%%%%%%%%%%%%%%%
\end{abstract}

%%%%%%%%%%%%%%%%%%%%%%%%%%%%%%%%%%%%%%%%%%%%%%%%%%%%%%%%%%%%%%%%%%%%%%%%%%%%%%%
%%%%%%%%%%%%%%%%%%%%%%%%%%%%%%%%%%%%%%%%%%%%%%%%%%%%%%%%%%%%%%%%%%%%%%%%%%%%%%%

\maketitle

%%%%%%%%%%%%%%%%%%%%%%%%%%%%%%%%%%%%%%%%%%%%%%%%%%%%%%%%%%%%%%%%%%%%%%%%%%%%%%%
\section{Introduction}\label{sec:introduction}
%%%%%%%%%%%%%%%%%%%%%%%%%%%%%%%%%%%%%%%%%%%%%%%%%%%%%%%%%%%%%%%%%%%%%%%%%%%%%%%

Metal nanowires have been the subject of many experimental and
theoretical studies\ \cite{Agrait03}. An important feature is the
quantization of motion of electrons because of spatial confinements.
In the linear regime, the Coulomb interaction among the electrons
are not important, and the transport properties can be well
described by Landauer formula\ \cite{Datta95} in the framework of
the free electron model where the transmission probability can be
calculated at equilibrium. Under a finite bias, however, the
scattering states of right- and left-moving electrons in a nanowire
are populated differently, even if there is no inelastic scattering
within the nanowire. An adequate treatment of the electron-electron
interactions is therefore crucial to correctly describe this
non-equilibrium electron distribution in the contact. Some studies
of cohesion\ \cite{Zagoskin98,Bogachek00} and transport\
\cite{Xu93,Pascual97,Zagoskin98,Garcia00,Bogachek97} in metal
nanowires at finite voltages using continuum models did not include
electron-electron interactions so that the calculated transport and
energetics depended separately on both the left and right chemical
potentials $\mu_+$ and $\mu_-$, thus violating the ``gauge
invariance'' condition: The calculated physical quantities should
depend only on the voltage $eV=\mu_+-\mu_-$, and should be invariant
under a global shift of the bias since the total charge is
conserved.\ \cite{Christen96} Other calculations have been made for
non-equilibrium metallic contacts including the electron-electron
interactions within the local-density approximation.\
\cite{Hakkinen98,Nakamura99,Yannouleas98} However, these
calculations utilized the canonical ensemble, which is not
appropriate for an open mesoscopic system. Finally, a
self-consistent formulation of transport and cohesion at finite bias
has been developed based on {\it ab initio} and tight-binding
calculations,\
\cite{Todorov00,Ventra01,Brandbyge02,Mehrez02,Mozos02} but {\em ab
initio} calculations can thus far only simulate small-size systems.

In this paper, we use the extended nanoscale mean-field electron
model developed in Ref.\ \onlinecite{Zhang04} to investigate the
single-particle and Coulomb interaction effects on the cohesion,
transport and magnetic properties of metal nanowires at finite
voltages. Since those quantities are response functions, they can be
used to characterize how the single-particle motion of electrons and
their Coulomb interactions respond to the corresponding external
forces. Those properties have been analyzed by Zagoskin\
\cite{Zagoskin98} and Bogachek {\it et. al}\ \cite{Bogachek00} using
a free-electron model which can only take into account the
single-particle effect. Here we emphasize the necessity of an
adequate treatment of electron-electron interactions not only to
satisfy the gauge invariance, but also  to show that the response of
the Coulomb interactions to the applied voltage and magnetic field
can give rise important effects on the cohesion, transport and
magnetic properties at finite voltages.

This paper is organized as follows: In Sec.\ \ref{sec:Model}, we
briefly introduce the extended nanoscale  mean-field electron model.
The details of this model are described in Ref.\
\onlinecite{Zhang04}. The the cohesion, differential conductance,
and magnetic properties are studied in sections\ \ref{sec:Transport}
and \ref{sec:Magnetic}. Sec.\ \ref{sec:conclusion} presents some
discussions and conclusion.

\section{Model}
\label{sec:Model}

Here we briefly introduce the extended nanoscale  mean-field
electron model. Details can be found in Ref.\ \onlinecite{Zhang04}.
In this paper, we only consider the temperature $T=0$. We consider a
cylindrical metallic mesoscopic conductor connected to two
reservoirs with respective chemical potentials $\mu_{+(-)}
=\mu+eV_{+(-)}$, where $\mu$ is the electron chemical potential in
the reservoirs at equilibrium, and $V_{+(-)}$ is the voltage at the
left (right) reservoir. While there is no general prescription for
constructing a free energy for such a system out of equilibrium, it
is possible to do so based on scattering theory if inelastic
scattering can be neglected, i.e. if the length $L$ of the wire
satisfies $L\ll L_{in}$. In that case the scattering states within
the wire populated by the left (right) reservoir form a subsystem in
equilibrium with that reservoir and the dissipation only takes place
within the reservoirs for the outgoing electrons. Using the
hard-wall boundary and treating the electron-electron interactions
in the Hartree approximation, one can define a non-equilibrium
grand-canonical potential $\Omega$ of the system at zero temperature
as \cite{Christen97,Zhang04}
\begin{equation}
\label{eq:omega_cy}
  \Omega_0[R_0,V,U] = -\frac{4\varepsilon_FL}{3\lambda_F}
    \sum_{\substack{\alpha=\pm \\ \nu}}
    \left(
     \frac{\varepsilon_\alpha-\varepsilon_{\nu }}{\varepsilon_F}\right)^{3/2}
    \!\!\!-N_+U,
\end{equation}
where $\varepsilon_F$ and $k_F$ are the Fermi energy and Fermi wave
vector, respectively, $\varepsilon_\alpha =\mu_\alpha-U$, and
$\varepsilon_\nu = \varepsilon_F \gamma_\nu^2/k_F^2R_0^2$ is the
$\nu$-th transverse eigenvalue of an electron in a cylindrical wire
with radius $R_0$,  with $\gamma_\nu$ the zeros of Bessel functions,
$U$ is the Hartree potential energy, which must be self-consistently
determined, and $N_+$ is the total number positive background
charges and is taken to be
\begin{equation}
\label{eq:N+}
N_+=\frac{k_F^3R^2_0L}{3\pi}-\frac{k_F^2R_0L}{4}
+\frac{k_FL}{3\pi}.
\end{equation}
The second term on the r.h.s of Eq.\ (\ref{eq:N+}) corresponds to
the well know surface correction in the free-electron model,\
\cite{Stafford99} which is essentially equivalent to placing the
hard-wall boundary at a distance $d=3\pi/8k_F$ outside the surface
of the metal wire.\ \cite{Lang73} The last term represent an
integrated curvature contribution.

Based on the assumptions of our model, the Hartree energy $U$
is constant along the wire and can be self-consistently
determined by the following charge neutrality condition at a given voltage $V$,
\begin{equation}\label{eq:neutrality}
  Q=\frac12e[N_-(\mu_+-U)+N_-(\mu_--U)]-eN_+=0,
\end{equation}
where
\begin{equation}
N_{-}(\mu_\pm-U)=\frac{2L}{\lambda_F\sqrt{\varepsilon_F}}
\sum_{\nu}(\mu_\pm-U-\varepsilon_\nu)^{1/2}
\end{equation}
is the number of right (left)-moving electrons in the cylindrical
wire up to energy $\mu_{+(-)}-U$. The summation is over all states
such that $\varepsilon_\nu < \mu_{+(-)}-U$. The uniformity of the
Hartree potential $U$ can be destroyed either by the realistic
atomic structure of wire (including impurities in the wire), which
can cause both elastic and inelastic scatterings, or the nonideal
couplings between the wire and the reservoirs, which induce
backscattering (see Ref.\ \onlinecite{Zhang04} for detailed
discussion). For our purposes in this paper, these effects are not
important. Equation (\ref{eq:neutrality}) gives a
relation\cite{Christen96}
\begin{equation}\label{eq:U0}
  U=U_s+\frac12(\mu_++\mu_-)-\varepsilon_F,
\end{equation}
where $U_s$ is calculated with a symmetric voltage drop
$V_+=-V_-=\frac12V$ between the two ends of the wire. Equation\
(\ref{eq:U0}) will guarantee that all physical quantities calculated
in the following are just a function of voltage $V$, and not of
$\mu_-$ and $\mu_+$ separately.

The current in the cylindrical wire, according to our assumptions, is given as
\begin{equation}
I(R_0,V,U)=\frac{2e}{h}\sum_\nu\int_{\mu_--U}^{\mu_+-U} dE\ \theta(E-\varepsilon_\nu),
\end{equation}
where $h$ is the Planck constant.

\begin{figure}[t]
  \includegraphics[width=1\columnwidth,angle=0]{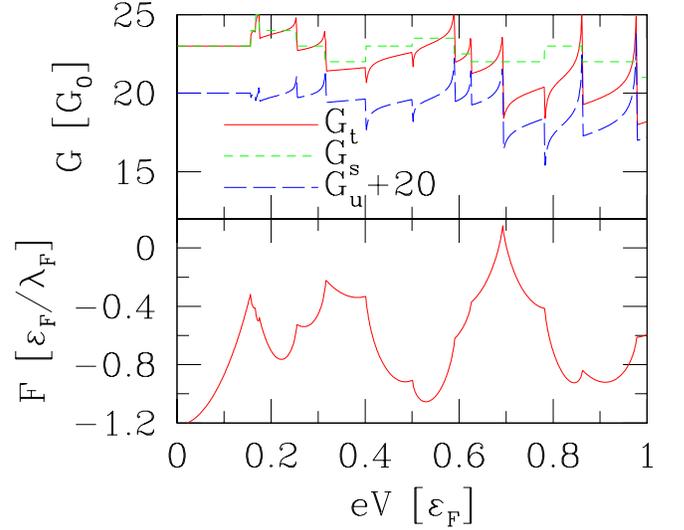}
  \caption{(Color line)The differential conductance and cohesive force of
a metal cylindrical nanowire versus voltage for a nanowire with
radius $k_FR_0=10.64$. Here $G_s$ and $G_u$ are the contributions
from single-particle motion and interaction effect, respectively,
and $G=G_t$ is the total differential conductance given by Eq.\
(\ref{eq:G_diff}).}
  \label{fig:gdiff}
\end{figure}
\begin{figure}[t]
  \includegraphics[width=1\columnwidth,angle=0]{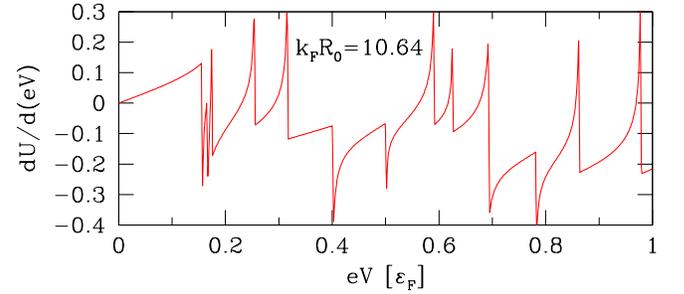}
  \caption{(Color line) The derivative of the Hartree energy $U$ with respect to the
           voltage.}
  \label{fig:dudvv}
\end{figure}
%

%%%%%%%%%%%%%%%%%%%%%%%%%%%%%%%%%%%%%%%%%%%%%%%%%%%%%%%%%%%%%%%%
\section{Cohesion and Differential Conductance}
\label{sec:Transport}
%%%%%%%%%%%%%%%%%%%%%%%%%%%%%%%%%%%%%%%%%%%%%%%%%%%%%%%%%%%%%%%%%%%%%%%%%%

The cohesive force $F =-\left.\frac{\partial\Omega[R_0,V,U]}
{\partial L}\right|_{N_+}$, where the derivative is taken at
constant background positive charge $N_+$, is given by
\begin{multline}
\label{eq:force}
F(R_0,V,U)= \frac{4\varepsilon_F}{3\lambda_F}\sum_{\substack{\alpha=\pm\\\nu}}
       \left(\frac{\varepsilon_\alpha-\varepsilon_\nu}
            {\varepsilon_F}\right)^{1/2} \\
    \times\left[\frac{\varepsilon_\alpha-\varepsilon_\nu}{\varepsilon_F}
    + 3k_FL\frac{\gamma_\nu^2}{k_F^3R_0^3}\left.\frac{dR_0}{dL}
\right|_{N_+}\right].
\end{multline}
The first term comes from the single-particle levels and the second
term is due to the finite size effect. The contribution from the
derivative of Hartree potential with respect to $L$ in the
electronic free energy  is completely canceled by that from the
positive charge background, and the interactions affect the cohesive
force only by shifting the chemical potentials, or equivalently by
shifting the single-particle levels by an amount of the Hartree
potential $U$. Therefore, the cohesion of metal nanowires is
determined by the single-particle motion of the electrons, even at
finite voltages.

The differential conductance $G= \frac{\partial I}{\partial V}$ is
given by
\begin{equation}
\label{eq:G_diff}
G(R_0,V,U) = \frac{G_0}{2}\sum_{\substack{\alpha=\pm\\\nu}}
      \theta\left(\varepsilon_\alpha-\varepsilon_\nu\right)
      \bigg[1- \alpha\frac{\partial U}{\partial(eV)}\bigg],
\end{equation}
where $G_0=2e^2/h$ is the unit quantum conductance. We see that both
single-particle motion and interactions contribute to the
differential conductance.

The differential conductance $G$ and cohesive force as a function of
voltage $V$ are shown in Fig.\ \ref{fig:gdiff}. In the figure, we
have split the total differential conductance $G$ into the
single-particle contribution part $G_s$ and the interaction
contribution part $G_u$.  At small voltages, $G=G_s$ and is equal to
that obtained from free electron model, which is consistent with the
result of linear transport. At large voltages, there are peak
structures at the conductance jumps, which come from $G_u$. This is
because at the subband thresholds, a small voltage change can induce
large fluctuation of the Hartree potential, as can be seen in Fig.\
\ref{fig:dudvv}. From Fig.\ \ref{fig:gdiff}, one can also see the
correlation between the variations of the conductance jumps and the
force oscillation as a function of the voltage, which should be
observable experimentally.

We should mention that an adequate treatment the electron-electron
interactions (Hartree approximation in this work), is essential to
satisfying the gauge invariance.
The calculated cohesive force and differential conductance
are just functions of the voltage between the two ends of the wire.
This is in contrast to the results
in Refs.\ \onlinecite{Zagoskin98} and \onlinecite{Bogachek00},
which depends on the partition $\beta$ of the voltage drops between the two contacts
by using a charge neutrality: $
N_-(\varepsilon_F+\beta eV)+N_-(\varepsilon_F-(1-\beta)eV)=2N_-(\varepsilon_F)$.
This treatment is clearly not self-consistent and does not satisfy the
condition of gauge invariance. We should also mention
that we should treat the system in the grand canonical ensemble so that
the Hartree potential on the force oscillation is not double-counted, as did
in Ref.\ \onlinecite{Ruitenbeck97}.

%%%%%%%%%%%%%%%%%%%%%%%%%%%%%%%%%%%%%%%%%%%%%%%%%%%%%%%%%%%%%%%%%%%%%%%%%%
\section{Magnetic Properties}
\label{sec:Magnetic}
%%%%%%%%%%%%%%%%%%%%%%%%%%%%%%%%%%%%%%%%%%%%%%%%%%%%%%%%%%%%%%%%%%%%%%%%%%

We have already seen that it is very important to treat the
electron-electron interactions appropriately to calculate the
cohesive force and the differential conductance. The interactions
can also produce significant effects in magnetic properties. By
using the free electron model, the calculated magnetoconductance
coefficient $\sigma$ and magnetotension coefficient $\Upsilon$ in
are identically zero\ \cite{Zagoskin98}. However, since the Hartree
potential $U$ has to be determined self-consistently by Eq.\
(\ref{eq:neutrality}) in presence of the magnetic field. As we will
show that, the response of Hartree potential is sensitive to the
external magnetic field at thresholds of the single-particle
subbands, and gives rise non-zero magnetoconductance coefficient and
magnetotension coefficient at these sub-bands. We should also show
that this response of Hartree potential to the magnetic field also
affects the magnetic susceptibility.

Using perturbation theory, the effect of a weak longitudinal
magnetic field $H$ (i.e. a field such that the cyclotron radius
$r_c\gg R_0$ perpendicular to the cross section of the wire) can be
included as a spin-dependent shift of the transverse
eigenvalues\cite{Zagoskin98}
\begin{equation}\label{eq:transverse}
  \varepsilon_{\nu s}(H) = \varepsilon_F\left[
    \frac{\gamma^2_\nu}{k_F^2R_0^2}+(m_{\nu}+gm_s)\frac{H}{H_0}\right],
\end{equation}
where $H_0=k_F^2hc/2e$, $g$ is the gyro magnetic ratio factor, $c$
is the speed of light, $m_\nu$ is the orbital angular momentum, and
$m_s=\pm\frac12$ is the spin of an electron.

\begin{figure}[t]
  \includegraphics[width=1\columnwidth,angle=0]{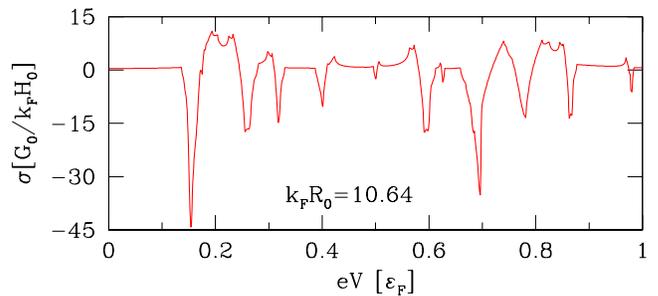}
  \caption{(Color line) The magnetoconductance coefficient $\sigma$ versus the
bias voltage.}
  \label{fig:sigma}
\end{figure}

\begin{figure}[t]
  \includegraphics[width=1\columnwidth]{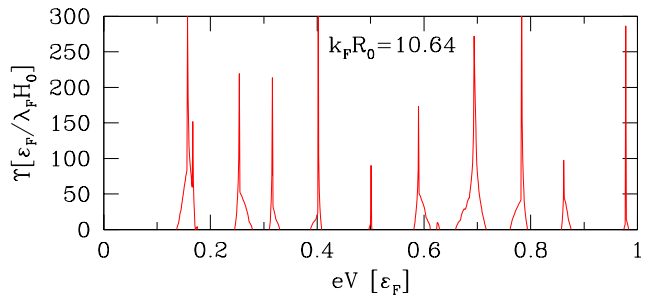}
  \caption{(Color line)
The magnetotension coefficient $\Upsilon$ versus the bias voltage.}
  \label{fig:upsilon}
\end{figure}

The generalized grand-canonical potential $\Omega$ and the current
at weak magnetic field are modified to be
\begin{equation}
\label{eq:omega_H}
  \Omega_0[R_0,V,H] = -\frac{2\varepsilon_FL}{3\lambda_F}
    \sum_{\substack{\alpha=\pm \\ \nu, s}}
    \left(
     \frac{\varepsilon_\alpha-\varepsilon_{\nu s}}{\varepsilon_F}\right)^{3/2}
    \!\!\!-N_+U,
\end{equation}
and
\begin{equation}
I(R_0,V,H)=\frac{2e}{h}\sum_{\nu,s}\int_{\mu_--U}^{\mu_+-U} dE\
\theta(E-\varepsilon_{\nu s}).
\end{equation}
The Hartree potential $U$ should be self-consistently determined by
Eq.\ (\ref{eq:neutrality}) under both voltage and the magnetic field.
The cohesive  force $F$ and differential conductance $G$ are also modified to
be
\begin{multline}
\label{eq:F_H}
  F(R_0,V,H) = \frac{2\varepsilon_F}{3\lambda_F}\sum_{\substack{\alpha=\pm\\
    \nu, s}}\left(\frac{\varepsilon_\alpha-\varepsilon_{\nu s}}
    {\varepsilon_F}\right)^{1/2} \\
  \times \left[
    \frac{\varepsilon_\alpha-\varepsilon_{\nu s}}{\varepsilon_F}
    + 3k_FL
      \frac{\gamma^2_\nu}{k_F^3R_0^3}
      \left.\frac{dR_0}{dL}\right|_{N_+}\right],
\end{multline}
and
\begin{equation}
\label{eq:G_H}
  G(R_0,V,H) = \frac{G_0}{4}\sum_{\substack{\alpha=\pm\\ \nu, s}}
    \theta\left(\varepsilon_\alpha-\varepsilon_{\nu s}\right)
  \left[1 - \alpha\frac{\partial U}{\partial(eV)}\right].
\end{equation}
The calculated cohesive force and differential conductance at small
magnetic field from Eqs.\ (\ref{eq:F_H}) and (\ref{eq:G_H}) are not
much different from Eqs.\ (\ref{eq:force}) and (\ref{eq:G_diff}).
These two equations serve the purpose to calculate the
magnetotension and magnetoconductance coefficients below.

The magnetotension coefficient $\Upsilon$ is defined as $\Upsilon=(\partial
F/\partial H)_{H=0}$ and magnetoconductance coefficient $\sigma$ is
defined as $\sigma=\frac{1}{L}(\partial G/\partial H)_{H=0}$.
Using Eqs. (\ref{eq:F_H}) and (\ref{eq:G_H}), one gets
\begin{align}
  \sigma(V) &= -\frac{G_0}{2} \sum_{\alpha=\pm}
    \left(1 - \alpha\frac{\partial U}{\partial(eV)}\right)
    \left.\frac{\partial U}{\partial H}\right|_{H=0}g(\varepsilon_{\alpha})
\nonumber\\
&\ \ \ \ -\frac{G_0}{2}\sum_{\substack{\alpha=\pm\\\nu}}
\alpha\theta(\varepsilon_\alpha-\varepsilon_\nu)
\left.\frac{\partial^2U}{\partial H\partial(eV)}\right|_{H=0},
\end{align}
and
\begin{multline}
  \Upsilon(V)= -\frac{2}{\lambda_F}
    \sum_{\substack{\alpha=\pm\\\nu}}\left[\left(
      \frac{\varepsilon_\alpha-\varepsilon_{\nu}} {\varepsilon_F}
      \right)^{-1/2}\frac{L\gamma^2_\nu}{k^2_FR^3_0}
\left.\frac{\partial R_0}{\partial L}\right|_{N_+}\right. \\
+\left.\left(\frac{\varepsilon_{\alpha}-\varepsilon_{\nu}}{\varepsilon_F}
\right)^{1/2}\right]\times \left.\frac{\partial U}{\partial H}\right|_{H=0},
\end{multline}
where $g(\varepsilon_{\alpha})$ is the density of states at energy
$\varepsilon_\alpha$ of electrons injected from reservoir $\alpha$.
From these two equation, one can see that nonzero magnetotension and
magnetoconductance coefficients are attributed to the response of
the Hartree potential $U$ to the magnetic field.

\begin{figure}[t]
  \includegraphics[width=1\columnwidth,angle=0]{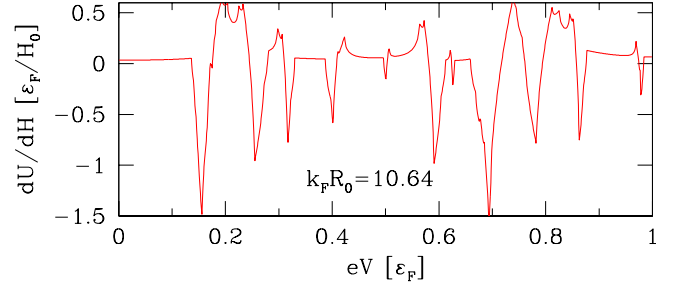}
  \caption{(Color line) The derivative of the Hartree energy $U$ with respect
to the magnetic field $H$ versus the bias voltage.}
  \label{fig:dudh}
\end{figure}

The calculated results for $\sigma$ and $\Upsilon$ are shown in
Figs.\ \ref{fig:sigma} and\ \ref{fig:upsilon}. Whenever there is a
subband opening, there is a peak in these two coefficients. This is
because the magnetic field increases the fluctuation of the Hartree
potential at those subbands, which can be seen in Fig.\
\ref{fig:dudh}.

\begin{figure}[t]
  \includegraphics[width=1\columnwidth,angle=0]{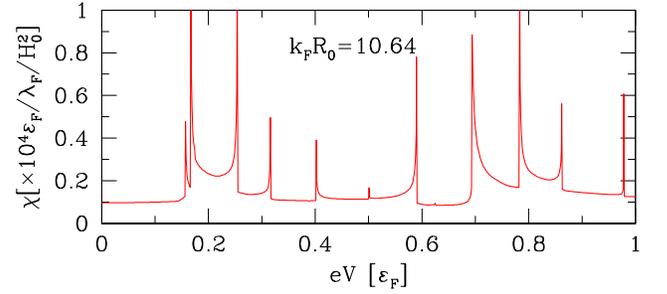}
  \caption{(Color line) The magnetic susceptibility $\chi$ versus voltage.}
  \label{fig:chi}
\end{figure}

One can consider the single-particle and interaction effects on the
magnetic susceptibility which is define as $\chi =
-\frac1L\left(\frac{\partial^2\Omega_0} {\partial
H^2}\right)_{H=0}$. Using the generalized grand canonical potential
Eq.\ (\ref{eq:omega_H}), and keeping only the dominant part, one
gets
\begin{multline}
  \chi(eV) \approx \frac{1}{2\varepsilon_F\lambda_F}
    \sum_{\substack{\alpha=\pm\\\nu s}}
    \left(\frac{\varepsilon_\alpha-\varepsilon_{\nu s}}
{\varepsilon_F}\right)^{-1/2}
\\
\times \left.\left( \frac{\partial(U+\varepsilon_{\nu s})}{\partial
H}\right)^2\right|_{H=0}.
\end{multline}
The result is presented in Fig.~\ref{fig:chi}. For nanowires with
small radii $R_0$, $\frac{\partial U}{\partial  H}$ is of the same
order as $\frac{\partial\varepsilon_{\nu s}}{\partial H}
=\varepsilon_F(m_\nu+gm_s)/H_0$,  both single-particle effect and
interaction effects are important to the magnetic susceptibility.
For nanowires with large $R_0$, $\frac{\partial U}{\partial  H}$ is
negligible  comparing to $\frac{\partial\varepsilon_{\nu
s}}{\partial H}$.

%%%%%%%%%%%%%%%%%%%%%%%%%%%%%%%%%%%%%%%%%%%%%%%%%%%%%%%%%%%%%%%%%%%%%%%%%%
\section{Discussions and Conclusion}
%%%%%%%%%%%%%%%%%%%%%%%%%%%%%%%%%%%%%%%%%%%%%%%%%%%%%%%%%%%%%%%%%%%%%%%%%%
\label{sec:conclusion}

We should point out that based on our recent stability analysis
\cite{Zhang04}, the nonzero $\sigma$ and $\Upsilon$ and the spikes
of $G$ and $\chi$ usually appear in the mechanical unstable zones.
However, this mechanical instability should  be a problem only in
the measurement of $\Upsilon$ since such a measurement is a
mechanical process. The sensitivity of the Hartree potential to the
applied voltage and magnetic field can be observed by measuring the
differential conductance $G$, magnetoconductance coefficients and
magnetic susceptibility as a function of the applied voltage as long
as the measuring time is shorter than the lifetime caused by the
mechanical instability.

In conclusion, we have used the generalized mean field electron
model at finite voltage bias\ \cite{Zhang04} to analyze the
single-particle and interaction effects on the cohesion, transport
and magnetic properties of cylindrical metal nanowires. At finite
voltage bias, it is crucial to treat the electron-electron
interactions adequately so that the calculated physical quantities
are gauge invariant. Our results show that the cohesive force is
determined by the single-particle effect, while  the nonzero
magnetotension and magnetoconductance coefficients are attribute of
the response of the Hartree potential to the magnetic field. Both
single-particle and interaction effects are important to the
differential conductance and the magnetic susceptibility.

\begin{acknowledgments}
This work was supported by NSF Grant Nos.\ 0312028 and 0454699. The
author thanks Charles A. Stafford, J\'er\^ome B\"urki and Herb
Fertig for useful discussions.
\end{acknowledgments}

\bibliography{zhang05}

\end{document}